# Localization of light by magnetically tuned correlated disorder : Trapping of light in ferrofluids

M. Shalini, Hema Ramachandran* and N. Kumar

Raman Research Institute, C.V. Raman Avenue, Sadashivanagar, Bangalore, INDIA 560080

Abstract

The mean free path of light ($l^*$) calculated for elastic scattering on a system of nanoparticles with spatially correlated disorder is found to have a minimum when the correlation length is of the order of the wavelength of light. For a typical choice of parameters for the scattering system, this minimum mean free path ($l^*_{\min}$) turns out to satisfy the Ioffe-Regel criterion for wave localization , $l^*_{\min} \sim \lambda$, over a range of the correlation length, defining thus a stop-band for light transmission. It also provides a semi-phenomenological explanation for several interesting findings reported recently on the transmission/ reflection and the trapping/storage of light in a magnetically tunable ferrofluidic system. The subtle effect of structural anisotropy, induced by the external magnetic field on the scattering by the medium, is briefly discussed in physical terms of the anisotropic Anderson localization.

## I. INTRODUCTION

Localization of light, or of a wave in general, in a strongly scattering disordered medium is now a well established phenomenon[1]. This arises from the interference of the coherently, multiply scattered waves in the random medium. Such a localization of a de Broglie wave was first studied theoretically by Anderson[2] in a seminal "oft-quoted but seldom read" paper for the case of an electron moving in a disordered lattice. He had predicted that for sufficiently strong disorder, electrons should become spatially localized turning the conductor into an insulator -- a phenomenon now referred to as Anderson transition. The corresponding case for light localization has been the subject of much theoretical[1,3] and experimental investigations[4,5] in recent years. Nominally, localization occurs when a certain general condition, namely, the Ioffe-Regel criterion is satisfied which for any classical or quantum wave, e.g., light or the de Broglie electron wave, requires that the transport mean free path $l^*$ be of the order of or less than the wavelength $\lambda_0$ of light in the medium. The Ioffe-Regel criterion essentially states that for the disorder - induced localization the mean free time ($t= l^*/v$, where v is the wave velocity in the medium) elapsed between the successive elastic scatterings should be less than the time period of the wave $2\pi/\omega$, i.e., $l^*/v \leq 2\pi/\omega$ , and hence $l^* \leq \lambda_0$. The exact numerical prefactor on the right hand side of the above inequality depends on the specific model of disorder chosen for the medium. Indeed, experimentally it has been reported[6] that at the onset of localization for an electron wave (of Fermi wavelength $\lambda$), $l^*/\lambda \approx 5.2$ implying $l^* \approx \lambda$. This criterion suggests that an increase in the wavelength, or a decrease in the transport mean free path should serve to localize light. The two quantities, however, are not independent, inasmuch as in the limit of long wavelength (i.e., in the Rayleigh limit), the scattering cross-section varies as the inverse fourth power of the wavelength $\lambda_0$ and the transport mean free path $l^*$ increases correspondingly. In the opposite limit of short wavelengths, on the other hand, one is in the ray-optical regime, where the wave interference is relatively less effective. While the optical analogue of a truly bound state (i.e., with negative energy) for an electron cannot obtain for light, it has been shown theoretically[3,7] that, similar to the case of sharp impurity states lying in the band-gap of a semiconductor, localized states of light can indeed occur within the energy gap of a photonic band-gap material upon the introduction of some defects (disorder) in the otherwise perfect periodic structure. Here localization is helped by the reduction in the photonic density of states in the gap (the Purcell-factor effect[7,8]). A direct manifestation of the light wave localization and of the thinning of the optical density of states is the well-known inhibition of spontaneous emission from an electronic impurity state lying localized within the gap[7,9]. Very recently, Schwartz *et al.*[10] have experimentally demonstrated the transverse localization of light caused by random fluctuations in a two-dimensional photonic crystal lattice.

In the present work, we re-visit the problem of light wave localization basing on the fact that the scattering cross-section for the wave has a pronounced maximum when the correlation length ($\xi$) of the correlated disorder in the medium matches the wavelength $\lambda$ $\lambda$/refractive index) of light in the medium. This should lead to a minimum in the transport mean free path ($l^*$) that may now well satisfy the Ioffe-Regel criterion, $l^* \sim \lambda_0$; a condition which is normally hard to satisfy for the light scattered by dielectric contrasts (the transport mean free paths usually obtainable through dielectric contrasts being typically 100μm to a few microns >> $\lambda_0$). We will later address the question of the tunability of the correlation length ξ that would give rise to a stop-band for the transmission of light, when we finally make contact with some recent experiments on the trapping of light in ferrofluids.

## II. DERIVATION

We begin by giving an analytical treatment for the scattering of light by a system of scatterers with correlated positional disorder. The transport mean free path ($l^*$) for the light wave, which in general involves the product of the form factor (namely, the individual scattering cross-section that characterises a single scatterer) and the structure factor (that takes into account the positional correlation in the distribution of the scatterers) is given by:

$$(l^*)^{-1} = \int \sigma\,(\theta,\phi)\; S(\mathbf{q})\,(1-\cos\theta)\; d\Omega \qquad \text{……….(1)}$$

with σ(θ,φ) the elastic differential scattering cross-section for a single scatterer (i.e., σ(θ,φ)$d\Omega$ is the flux scattered into the solid angle element $d\Omega$ about the scattering direction (θ,φ) for a unit incident flux). Here $\mathbf{q} = \mathbf{k_s} - \mathbf{k_i}$ is scattering wave vector (the momentum transfer divided by ħ), with $|\mathbf{k_s}| = |\mathbf{k_i}|$ and the magnitude q = 2k sin(θ/2); $\mathbf{k_i}$ and $\mathbf{k_s}$ being, respectively, the incident and the scattered wave-vectors. The structure factor is given by

$$S(\mathbf{q}) = \frac{1}{V}\left\langle \sum_{i,j} e^{-i\mathbf{q}.(\mathbf{R_i}-\mathbf{R_j})} \right\rangle \qquad \text{……(2)}$$

with $\mathbf{R}_i$ the position of the $i^{\text{th}}$ scatterer. The factor (1-cos θ) weights the effectiveness of the backscattering in calculating the transport mean free path $l^*$. Thus, for example, for a totally random collection (dilute gas, say) of point-like scatterers (of size $a_0 << \lambda_0$) we have the Rayleigh scattering off the individual scatterers with a total single particle scattering cross-section σ given by

$$\sigma = \int_0^{\pi} \sigma(\theta,\varphi)\; 2\pi \sin\theta\; d\theta = \frac{8\pi}{3}\, k^4\, a_0^{\,6} \left| \frac{(\varepsilon-1)}{(\varepsilon+2)} \right|^2 \qquad \text{………(3)}$$

Here ε is the dielectric constant of the material of which the scatterers are made relative to the dielectric constant of the background. The total scattering cross-section for the N random scatterers is then simply Nσ. Moreover, in this case the structure factor is trivial :

$$S(\mathbf{q}) = \frac{1}{V}\left\langle \sum_{i,j} e^{-i\mathbf{q}.(\mathbf{R_i}-\mathbf{R_j})} \right\rangle$$

$$= \frac{N}{V}\left[1 + \delta_{q,0}\,(N-1)\right]$$

$$= \mathrm{n}[1 + \delta_{\mathrm{q},0}\,(\mathrm{N}-1)] \qquad \text{........(4)}$$

where n = N/V, is the mean number density of the scatterers. Thus, for this case of the totally uncorrelated random gas of Rayleigh scatterers, we have the well-known expression for the transport mean free path $l^*$:

$$l^* = \frac{1}{(n\sigma)} \qquad \text{........(5)}$$

We will now consider such an isotropic system of scatterers, but with positional correlation described by a non-trivial structure factor :

$$= \frac{1}{V}\iint \sum_{i,j} e^{-i\mathbf{q}.(\mathbf{r}-\mathbf{r}')} \left\langle \delta(\mathbf{r}-\mathbf{R_i})\,\delta(\mathbf{r}'-\mathbf{R_j}) \right\rangle d\mathbf{r}\,d\mathbf{r}'$$

$$= \frac{1}{V}\iint \sum_{i,j} e^{-i\mathbf{q}.(\mathbf{r}-\boldsymbol{\rho}')} \left\langle n(\mathbf{r})\,n(\mathbf{r}') \right\rangle d\mathbf{r}\,d\mathbf{r}' \qquad \text{........(6)}$$

With this, we can re-write the transport mean free path as

$$(l^*)^{-1} = \frac{1}{V}\iiint \sigma(\theta,\varphi)(1-\cos\theta)\; e^{-i\mathbf{q}.(\mathbf{r}-\mathbf{r}')} \left\langle n(\mathbf{r})\; n(\mathbf{r}') \right\rangle\; d\Omega\;\; d\mathbf{r}\,d\mathbf{r}' \qquad \text{....(7)}$$

Now, the spatial correlation of the density fluctuations can be introduced and characterized generally through a correlation length scale ξ. For our semi-phenomenological treatment, we will take the correlation to have an exponential spatial decay. With this, we obtain

$$\left\langle n(\mathbf{r})\;\, n(\mathbf{r}') \right\rangle = \left\langle \delta n(\mathbf{r})\,\delta n(\mathbf{r}') \right\rangle + n_0^{\;2}$$

$$\approx \left\langle \delta n^2(0) \right\rangle e^{-|\mathbf{r}-\mathbf{r}'|/\xi} + \mathrm{K} \qquad \text{………(8)}$$

where, the density fluctuation $\delta(n) = n(\mathbf{r}) - n_0$ with $n_0$ the uniform mean value, and the ellipses denote the trivial uniform background that clearly does not scatter. The statistical fluctuation $\langle \delta n^2(0) \rangle$ above is to be taken as an overall prefactor without any sensitive dependence on the degree of order. Thus, e.g., for the Poissonian fluctuations (as will be assumed here for simplicity) within the interparticle volume $\sim 1/n_0$, we have $\langle \delta n^2(0) \rangle = n_0^2$. With this proviso, we obtain for the transport mean free path -

$$(l^*)^{-1} = n_0^{\,2} \iint \sigma(\theta,\varphi)(1-\cos\theta)\; e^{-i\mathbf{q}.\mathbf{r}}\; e^{-r/\xi}\, d\mathbf{r} \quad d\Omega \qquad \text{………… (9)}$$

For the case of our isotropic (nanometric-scale spherical) scatterers, we have $\sigma(\theta,\varphi)=\sigma/4\pi$. Further, an important and well known fact[11] is that a collection of $n_0\lambda_0^3$ scatterers lying within an optical cell volume $\sim\lambda_0^3$ scatters phase-coherently making the effective scattering cross-section for the cell $(n_0\lambda_0^3)^2\sigma$. (We recall that $\sigma$ here is the total scattering cross-section for a single scatterer). Thus, for the case of our rather dense random scatterers, we must modify the above expression by multiplying the single particle differential scattering cross-section $\sigma(\theta,\varphi)$ by the coherence enhancement factor $n_0\lambda_0^3$ . (By way of contrast, we should note that for the electronic case this coherence factor is essentially ~1 because of the smallness of the electron de-Broglie (Fermi) wavelength). Accordingly, on carrying out the space integral over $\mathbf{r}$, we obtain for the normalized transport mean free path,

$$\left(\frac{l^*}{\lambda_0}\right)^{-1} = \left(\frac{256\pi^6}{3}\right) n_0^{\,3}\, a_0^{\,6}\, \lambda_0^{\,3} \left|\frac{(\varepsilon-1)}{(\varepsilon+2)}\right|^2 \int_0^{\pi} \frac{\sin\theta(1-\cos\theta)\left(\dfrac{\xi}{\lambda_0}\right)^3}{\left[1+16\pi^2\left(\dfrac{\xi}{\lambda_0}\right)^2 \sin^2\left(\dfrac{\theta}{2}\right)\right]^2} d\theta\,, \qquad \text{………(10)}$$

where, we have used $\displaystyle \int e^{-i\mathbf{q}.\mathbf{r}}\; e^{-r/\xi}\, d\mathbf{r} = \frac{8\pi\xi^3}{\left[1+q^2\xi^2\right]^2}$

Recall that $q = (4\pi/\lambda_0)\sin(\theta/2)$, with $\theta$ the scattering angle. Performing the integral over $\theta$, we finally obtain –

$$\left(\frac{l^*}{\lambda_0}\right)^{-1} = \frac{16\pi^6 n_0^{\;3} a_0^{\;6} \lambda_0^{\;3}}{3} \left|\frac{(\varepsilon-1)}{(\varepsilon+2)}\right|^2 F(X), \qquad \text{..........(11)}$$

where

$$F(X) = \frac{[1+16\pi^2 X^2]\ln[1+16\pi^2 X^2] - X}{X^3[1+16\pi^2 X^2]} \qquad \text{.........(12)}$$

with $X = \xi/\lambda_0$. Note that F(X) is a universal function. Equation(11) is the main result of our derivation.

A few comments are now in order for the enhancement of scattering when the wavelength $\lambda_0$ matches the correlation length $\xi$. This is physically analogous to the well known phenomenon of critical opalescence, where, of course, the correlation length is tuned thermally as we approach the critical temperature, and becomes comparable to the wavelength of the light. Indeed, it is known that the light scattering (opalescence) is at its maximum at a temperature at which the correlation length equals the wavelength of light[12]. In the present context, the temperature tuning of the critical opalescence has been replaced by the magnetostatic tuning of the spatial correlation of the light scatterers. Here the field-induced ordering of the disordered ferrofluidic system has been parameterized phenomenologically through a correlation length, $\xi$, which is expected to increase monotonically with increasing external magnetic field which is responsible for inducing the spatial order. (It should be clarified, however, that we are not talking here about the spontaneous order and its correlation length as for a thermodynamic system approaching the critical point of a phase transition. In the present case, the forced order is created primarily from the consideration of magnetostatic energy minimization[13]). Thus, as the magnetic field is tuned up from the zero value, the spatial order and the associated correlation length must increase, and can become comparable to the wavelength of light, which should result in strong scattering. It is this strong scattering over a certain stop-band of the magnetic field that can enable us to satisfy the Ioffe-Regel criterion for the localization of light, that would otherwise demand a very strong dielectric contrast between the scatterers and the background medium. This would be hard to realize. Our phenomenological model based on localization in terms of tunable correlation length may be generally viewed as interpolating between complete randomness and complete order (photonic band-gap crystal). It may be further refined so as to take into account a more realistic model of magnetic ordering of the ferrofluidic system and the resulting anisotropy.

## III. RESULTS AND DISCUSSION

Figure 1 shows a plot of $l^*/\lambda_0$ ( transport mean free path normalised by the wavelength of light in the medium) against $\xi/\lambda_0$ (the correlation length normalised by the wavelength) obtained using Eq. 11, for a choice of parameters typical of the nanometric system to be discussed shortly. (The plot is for $\lambda_0$ = $\lambda$/refractive index of medium = (0.632/1.45) μm ; $a_0$, the size of the scatterers = $2\times10^{-2}$μm, $n_0 = 2\times10^3 \mu m^{-3}$, and $|(\varepsilon-1)/(\varepsilon+2)|^2=1/2$). Figure 1 clearly shows that the calculated transport meanfree path ($l^*$) dips below the wavelength value $\lambda_0$ for a range of the correlation length, $0.065 < \xi/\lambda_0 < 3.57$. In this range (shown as the shaded region in the graph), therefore, the Ioffe-Regel localization criterion is well satisfied, defining thereby a stop-band for the transmission of light through this optical medium. (It is apt to note here that while the criterion for the onset of localization is framed in terms of $l^*$ calculated in the first Born approximation, the resulting localization develops from the coherent multiple back-scattering which is a non-perturbative effect and involves $l^*$ as a parameter).

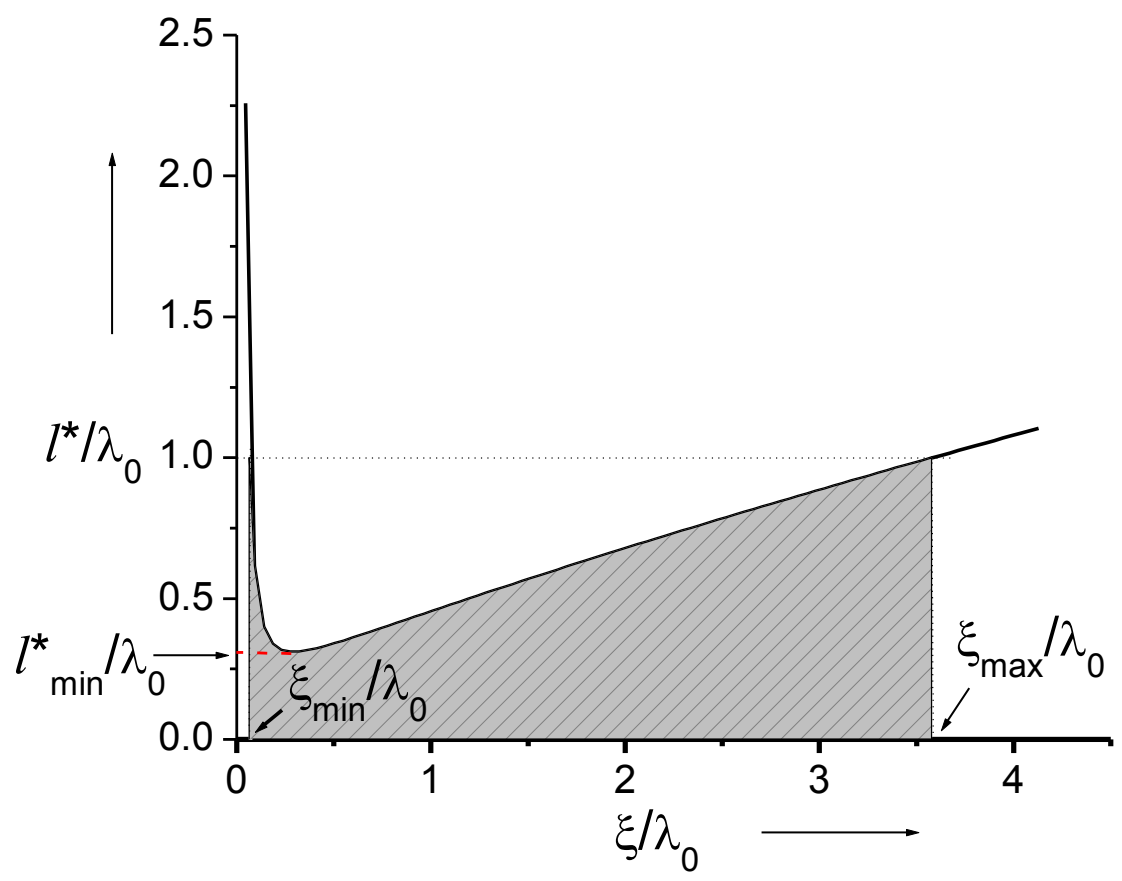

FIG. 1 : Plot of normalized mean free path ($l^*/\lambda_0$) against the normalized correlation length ($\xi/\lambda_0$) for a certain choice of parameters given in the text. The Ioffe-Regel criterion is satisfied in the shaded region; hence, the stop-band for light transmission.

In Fig.2, we have plotted the universal function F(X) (Eq. 12) for the sake of completeness.

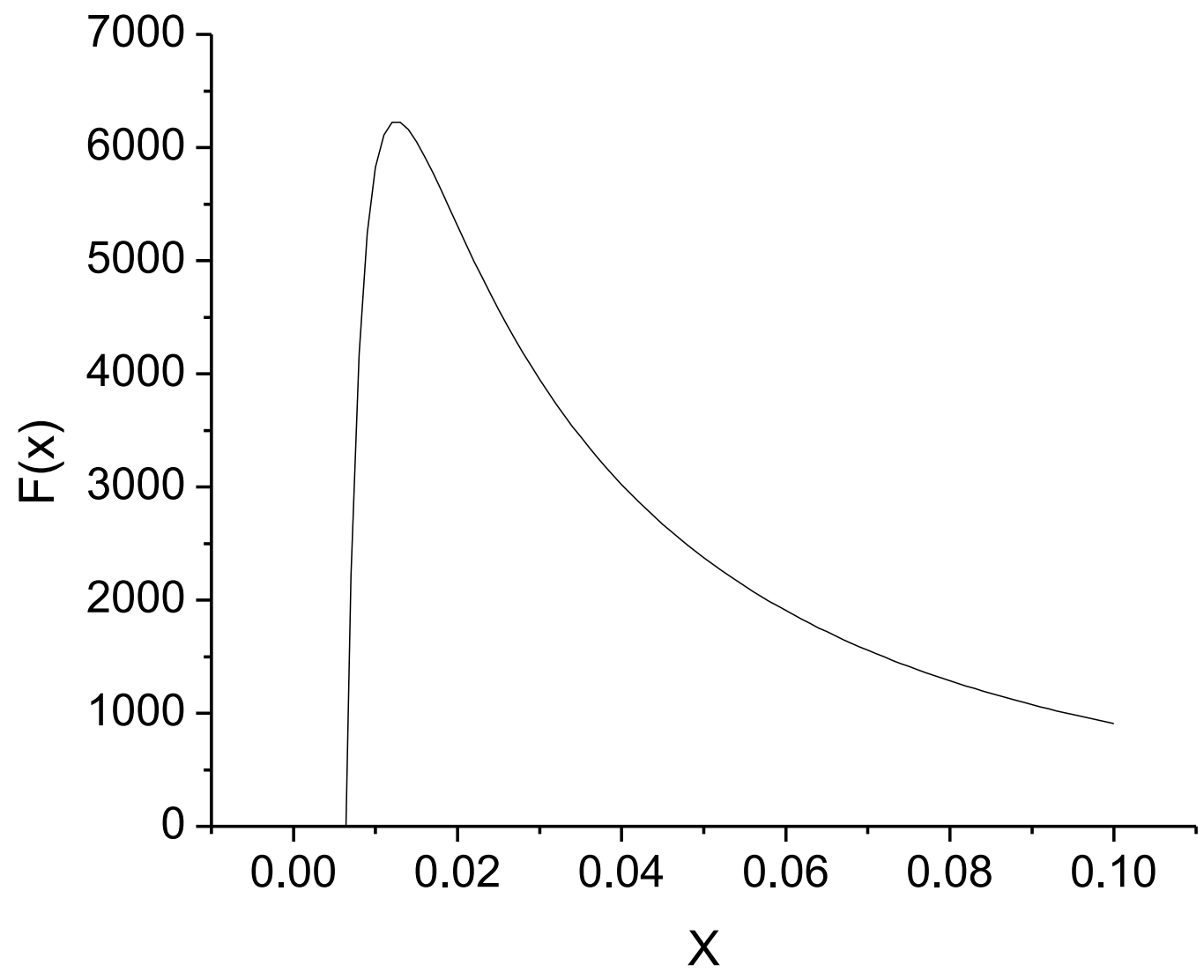

FIG.2: The universal function F(X) versus X

We will now attempt an interpretation of the recent experimental observations of Mehta and co-workers[14-17] in on the transmission/reflection and trapping/storage of light in the magnetically tunable ferrofluidic system in the light of our semi-phenomenological treatment presented above. These authors have reported two interesting, and possibly important experimental observations in a system consisting of a random suspension of nanospheres and microspheres of $Fe_3O_4$ (magnetite) in kerosene :

(A) Reversible blocking of light transmission through the disordered ferrofluidic scattering system over a stop-band of an external magnetic field applied transverse to the direction of the incident light, and the subsequent delayed (upto ~100ms) release of the stored light in the form of a flash upon removal of the magnetic field.

(B) Simultaneous vanishing of the transmitted and the reflected light on approaching the stop-band.

While these experimental observations are admittedly very interesting, the explanation for the observed phenomena offered[14] in terms of the vanishing of the forward as well as the backward scattering of light for a certain combination of the magnetic permeability and the dielectric permittivity (both reckoned at optical frequencies) of the spherical magnetite scatterers is basically flawed[18,19]. This is because at optical frequencies, the magnetic permeability $\mu$ =1 and there is no magnetic scattering to interfere destructively with the dielectric scattering.

Our explanation for these findings is rather straightforward when viewed in terms of localization of light in this remarkable disordered medium of random scatterers, where the spatial order can be induced and tuned magnetostatically by the externally applied magnetic field. It is well-known that the magnetite particles re-arrange to form linear chains in the direction of the applied magnetic field[20,21].

Let us first consider the observation (A). By varying the external magnetic field one may tune continuously the spatial correlation length, $\xi$, sweeping across the matching condition $\xi\sim\lambda_0$ giving a strong scattering resulting in localization as has been shown in our semi-phenomenological treatment above. This should naturally give a stop-band (bracketing the $\xi\sim\lambda_0$ condition) as has indeed been reported. Thus we believe that in the experiments of Mehta *et al.*, in the stop-band the light is stored or trapped *as light localized within the medium*. In this picture, we should now expect the stored light to exit as a flash as the magnetic field is switching off, bringing the system out of the stop-band. Indeed, the storage of light *as light* is consistent with their explicit demonstration of the pumping of a dye fluorescence by the trapped light[17].

We re-emphasize that all scattering here is essentially electric-dipolar in nature; the magnetic nature of the particles, however, helps order the scatterers in space magnetostatically through an externally applied magnetic field. It is this synergy between the electric-dipolar light scattering by the nano-scatterers and the magnetostatic tunability of their spatial order that makes the ferrofluidic system remarkable as a magnetically tunable scattering optical medium.

It is apt at this stage to distinguish between the properties of the ferrofluid (nanoparticles treated as Rayleigh scatterers) and those of the microspheres (treated as Mie scatterers) suspended in it, and also their relative roles in the localization of light. (Recall that typically $a < \lambda/10$ for Rayleigh scattering, $\lambda/10 < a < 10\lambda$ for Mie scattering, and $a > 10\lambda$ for the refraction/ reflection description to hold). In the experiment[14,15] the typical size of the magnetite ($Fe_3O_4$) nanospheres was about 10nm, and their number density was about $10^{15}cm^{-3}$. The magnetite microspheres, on the other hand, were much larger (2-3μm) in size and had a lower number density ($10^8$ $cm^{-3}$). The wavelength λ of light used was 0.632μm in vacuum, which gives $\lambda_0$ = 0.43μm in the medium (kerosene). The nanoparticles being of a small size, below the Marshall limit, are expected to be mono-domain[22], or subdomain, and thus act as permanent finite magnetic dipoles that can be ordered orientationally by an externally applied magnetic field. Their permeability at optical frequencies, however, stays essentially equal to 1 as noted earlier, and thus there is no magnetic dipolar scattering. The microspheres are too large (multi-domain) to have a net spontaneous magnetic moment, but are magnetostatically polarizable directly by an external magnetic field as also by the nanomagnets mediating between them. The microspheres too scatter light, but again primarily through their dielectric polarizability, and predominantly in the forward direction, which is not effective for localization. (Note the explicit (1-cosθ) factor in Eq.(1) for the effective scattering cross-section).

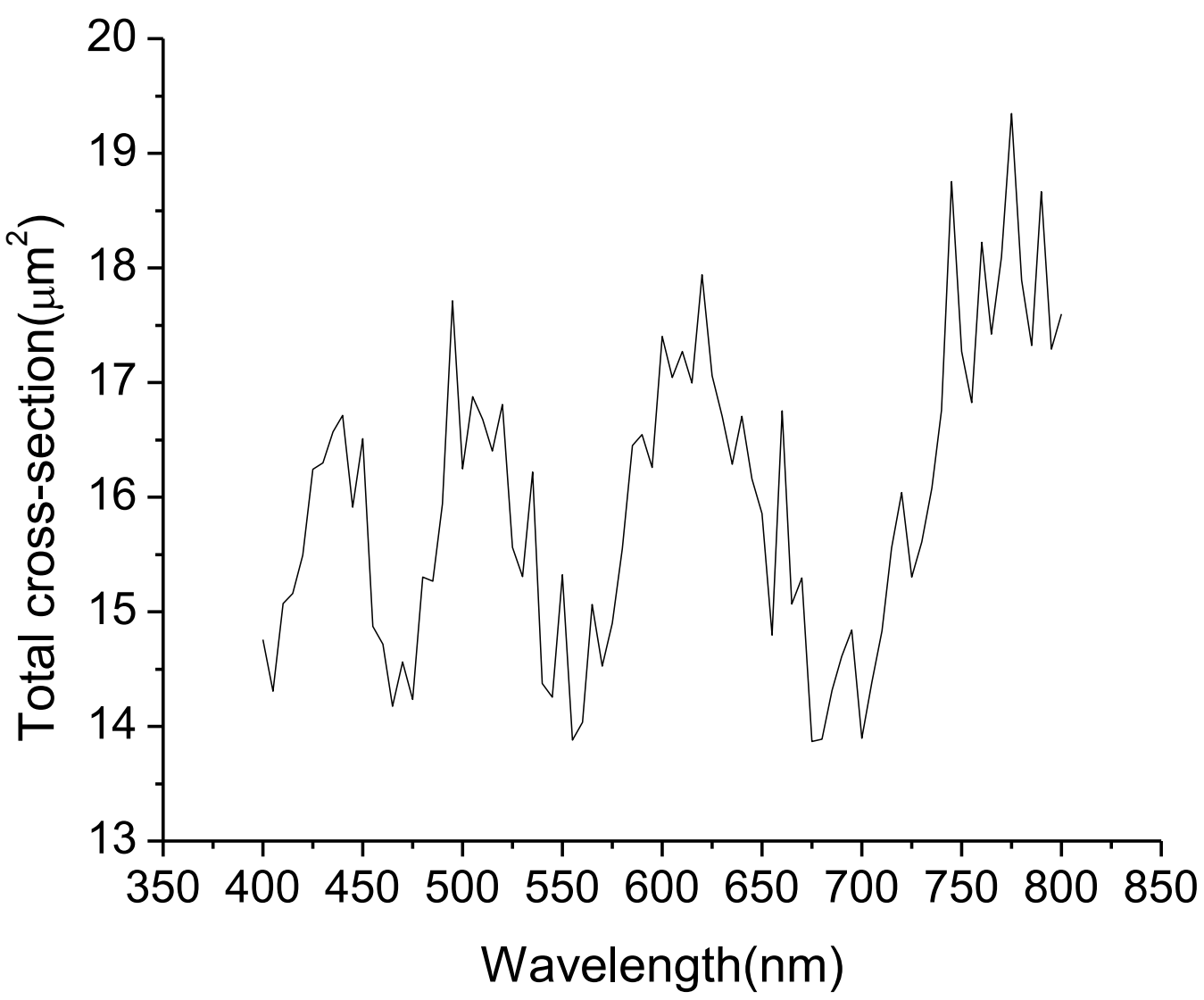

FIG.3 :. Calculated Mie scattering cross-section of the microspheres as a function of wavelength for parameters given in the text.

That the microspheres play a relatively sub-dominant role in light scattering in this system is borne out by our calculation of the transport mean free path[23,24]. The calculated total scattering cross-section of a single Mie scatterer is plotted in Fig.3 as a function of wavelength, for a choice of parameters corresponding to the experiment. Clearly, one sees peaks corresponding to the various multipole resonances, in fact with a pronounced peak at the wavelength used in the experiment. The transport mean free path calculated using Eq.7 for Mie scatterers, as in the case of the Rayleigh scatterers, also has a minimum as function of the correlation length; however, the minimum was relatively shallow. The role of the microspheres appears to us, in the present experiment, as essentially providing for a scaffolding for spatial magnetic order in the medium giving a finite correlation length as function of the externally applied magnetic field. Thus we believe that the light scattering is mainly by the nanoparticles. In addition, these magnetically orientable nanoparticles mediate interaction (bonding chains, say) between the polarizable microspheres, and help create the spatial order which is tunable magnetostatically externally. Magnetic ordering of similar systems has been studied by several workers[13, 20,21].

Now we turn our attention to observation (B), namely, the simultaneous vanishing of the transmitted and the reflected light on approaching localization (the stop-band). While away from localization, the transmitted light intensity is expected to be much greater that the reflected intensity as clearly observed experimentally, the opposite is expected to be the case as we approach the localization regime (the stop-band). It indeed then comes as a surprise that, experimentally, *both the intensities are actually found to diminish greatly and simultaneously* ! Where does the light go ? The question may be answered in physical terms of the anisotropic Anderson localization. Here it is known that despite the anisotropy, the "mobility edge" (the onset condition for localization) itself is independent of the direction, as in the case of isotropic Anderson localization. But, the associated length scales, namely, the wave-correlation length $L_{wc}$ (localization length $L_{loc}$) in the extended-state regime (localized-state regime) are NOT [25-27]. In particular, the wave-correlation length $L_{wc}$ along the direction of the relatively weaker disorder can be much (exponentially) smaller than its value in the direction of the relatively stronger disorder. Now, in the present case, the structural anisotropy of disorder is caused by the externally applied magnetic field which is known to create chain-like ordering of the scatterers along the direction of the magnetic field, i.e., transverse to the direction of the incident light (longitudinal direction). This partial ordering in the transverse direction implies a relatively much weaker disorder, and correspondingly a much smaller transverse wave-correlation length scale $L_{wc.}$. This in turns implies much greater transverse diffusion relative to that in the longitudinal direction as the stop-band is approached. We, therefore, expect the light to diffuse away sideways on approaching the stop-band, thus *escaping out in the transverse direction*. Clearly, this dominant transverse escape implies simultaneous depletion of the transmitted and the reflected intensity along the longitudinal direction. This is indeed what is observed experimentally.

Finally, we would like to comment briefly on a possible purely electronic mechanism for the phenomenon in question, namely that of storage/ trapping of light as an electronic excitation/resonance, and its subsequent delayed re-emission as a pulse, e.g., by fluorescence/phosphorescence as for instance in "quantum-dot blinking". However, for the cw incident light, the emission of stored/ trapped light energy as magnetically gated light pulse (flash) will again involve localization, in that it is the latter that inhibits the emission of light by de-excitation in the stop-band due to the absence of/ reduction in the propagating modes in the localization regime. Moreover, there is no experimental evidence for appreciable absorption in the optical band – the absorption band lies in the ultraviolet[28]. Besides, the fluorescence / phosphorescence does not preserve the polarization, while experimentally, the emerging flash of light is found to have the same polarization as the incident light. The multiple scattering in Anderson localization is, however, dominated by coherent back-scattering that does retain the polarization. We, therefore, believe that light energy is stored as light through Anderson localization by disorder which is magnetically tunable in the present case.

## IV. CONCLUDING REMARKS

We have provided a semi-phenomenological theory for the localization of light in a medium with magnetically tunable correlated disorder. The crucial point here is that the scattering is strongest for the correlation length matching the wavelength of light, for which an approximate analytical expression has been derived. This provides a qualitative understanding of the experimental findings of the reversible trapping/storage of light in a ferrofluidic system over a stop-band of an externally applied magnetic field. Our analytical treatment is based on an isotropic Anderson model, while, actually, the system is rendered anisotropic by the externally applied transverse magnetic field. This enables us to understand another important feature, namely, the simultaneous reduction of the transmission as well as the reflection as we approach the stop-band. Besides, it is known that anisotropy favours localization in that it effectively lowers the dimensionality of the system[26]. We hope that our work will encourage further experimental and theoretical investigations that should sharpen our understanding of these phenomena.

Acknowledgements : Authors would like to thank R.V. Mehta and R. Patel for sharing with them details of their early experimental findings, and also for participatory demonstration and measurements of some of these effects in the authors' laboratory.